\documentstyle[12pt]{article}
\begin{document}
\title{{\bf MINDLESS SENSATIONALISM:
\\A QUANTUM FRAMEWORK FOR CONSCIOUSNESS}
\thanks{Alberta-Thy-04-01;
invited contribution to be published in
{\em Consciousness:  New Philosophical Essays},
edited by Quentin Smith and Alexandar Jokic
(Oxford: Oxford University Press, 2002).
}}
\author{
Don N. Page
\thanks{Internet address:
don@phys.ualberta.ca}
\\
CIAR Cosmology Program, Institute for Theoretical Physics\\
Department of Physics, University of Alberta\\
Edmonton, Alberta, Canada T6G 2J1
}
\date{(2001 Feb. 15)}

\maketitle
\large
\normalsize
\baselineskip 15 pt

	I write as a physicist, with the knowledge that
many physicists are studying the mathematical structure
of our physical universe and hope to be able to find
a `theory of everything,' or TOE, that will give a complete
set of dynamical laws for this structure.
This would essentially give the rules for
how all the physical quantities in the universe evolve.
Such a TOE would not be itself give
the boundary conditions also necessary to determine
the history of the universe, so it
is a misnomer to say that it is a `theory of everything,'
but it is convenient to retain the acronym TOE
for this dynamical part of the laws of physics.
Some physicists, such as Hartle and Hawking
\cite{HarHaw},
are also seeking to find rules specifying
the boundary conditions (BC) of the universe as well.
The combination of the TOE and the BC
would then give a complete description
of the mathematics of the state of the universe and its evolution.
(This might be called a `Theory Of More of Everything,'
or TOME.)

	However, if one takes an even broader view,
one realizes that even the TOME (the TOE and the BC)
would not really comprise a theory of everything either,
since they do not specify what conscious experiences
occur within the universe.
At least this seems to be the case if
the TOME is assumed
to be of the general mathematical types
that are currently being sought,
since such types do not seem by themselves
to specify precisely what conscious experiences occur.

	Nevertheless, there is the general consensus
that there should be some sort of
`psycho-physical parallelism'
or connection between the mathematical structures
described by current and sought-for theories of physics
and the conscious experiences that each of us apparently has.
Indeed, it can be argued that all we directly
experience are these conscious experiences themselves,
and our feelings that there is a mathematical
structure for the physical world
seems to be ultimately based upon the enormous success
of our partial glimmerings of such a structure
in explaining many aspects of our conscious experiences.
In other words, we do not seem to experience directly
the mathematical structure at all,
but we seem to experience the feeling
that our partial theories for such a structure
help us better understand our experiences.

	For example, as part of some of my conscious
experiences while writing this, I have a feeling
that I am looking at a computer screen that
(except for the details of what is displayed upon it)
is very similar to what I would consider to be the same computer screen
that I think I remember viewing at many times in the past.
Furthermore, I have the feeling that my understanding
of the feeling of the existence and persistence of certain properties
of what I interpret to be the computer screen in front of me,
is helped by my effective partial theory of the
existence of this screen as a physical object and of its 
approximate ``object permanence'' over the relevant timescales.
(Incidentally, I do not believe that any ultimate theory
of physics will have any precisely existing persons such as ``I,''
any precisely existing ``objects'' such as computers screens,
any absolute notions of ``personal identity'' or ``object permanence,''
or even any absolute notion of time or of timescales,
but to illustrate my ideas,
I am merely using the crude notions from a rough
instrumentalist theory to denote how ``I''
feel ``I'' ``believe'' ideas about
an ``external'' ``physical'' world seem to help
explain my ``internal'' ``mental'' experiences.)

	Therefore, very crudely, I think that
I have the experience of remembering my computer screen
as a persisting object because, according to my rough theory,
there is such an object in the physical world.

	Such a rough theory can be refined,
and I might believe that a better theory claims that
my conscious experience is more directly correlated with
(or is ``caused by'') certain physical processes within my brain.
The point is that it certainly seems to have explanatory value
to assume that in some sense there exists a physical world,
and that our conscious experiences are either part of it
or else are correlated with it.

	Of course, it is logically possible that only the conscious
experiences by themselves exist (or even just the one
conscious experience that I am having ``now,'' to take an extreme
solipsistic view that denies even the existence
of my past experiences as anything other than the partial contents
within the memory components of my present experience).
However, the experienced correlation between the different components
of the content of even my present experience
would then seem to lack the explanation that appears possible
from the assumed existence of an external physical world.

	Therefore, I shall assume that an external physical world
does exist in some sense and is helpful for explaining
our conscious experiences.
For it to be helpful, it must be connected or correlated
with the internal conscious experiences in some way,
and this connection is the `psycho-physical parallelism' or PPP
that I shall assume exists.

	Now the question arises as to what the form is of this
assumed `psycho-physical parallelism.'
This form will of course be dependent upon the form
of the two entities that are being connected,
the internal conscious experiences and the external physical world.
I am certainly no expert on the academic work
that has been done on theories of the form of the internal conscious
experiences, though I can claim to experience at least one of them
directly myself.
On the other hand, I have done academic work for many years
on theories for the form of the external physical world,
and so I have some idea of the constraints of current
physical theories on that end of the psycho-physical connection,
even though we physicists certainly do not yet have the complete
physical theory or TOME described above,
and I would not be personally competent
to assess it fully even if we physicists as a community
did have such a theory.

	An essential point here is that, so far as we know,
and so far as current physics theories give any strong hint,
the external physical world seems to be thoroughly
quantum mechanical.
Therefore, as Quentin Smith
\cite{Smith}
has emphasized in this volume,
a realistic theory of the `psycho-physical parallelism' should
include the quantum nature of the physical world
in order to be consistent with the most basic feature
of our current best theories of physics.

	I should say that, unlike some, I do not believe
that it is necessarily impossible for there to exist
a (different) universe in which the physics is entirely classical
and yet conscious experiences exist and are correlated with
that external physical world.
However, I am strongly convinced that such a classical universe
is not ours, and so if we want a correct theory
of the psycho-physical connection for our universe,
we must include the quantum nature of our universe
(or possibly whatever it is that replaces the quantum
if our current quantum theories are entirely superseded,
though I think it highly unlikely that such a future theory
would revert entirely to the completely classical picture
held before quantum theory was discovered).

	Of course, there are a multitude of ways
in which one might postulate a connection between
conscious experiences and a quantum physical world.
Quentin Smith
\cite{Smith},
Barry Loewer
\cite{Loewer},
and Michael Lockwood
\cite{Lockwood}
have discussed three within this volume.
However, rather than reviewing the various possibilities
that have been proposed, I wish to summarize
my own conjecture for the framework or basic form
of the connection.
When emphasizing the quantum side of the connection,
I have called this Sensible Quantum Mechanics (SQM)
\cite{SQM,Page},
but, for reasons that will become apparent,
when emphasizing the conscious side of the connection,
I might call it Mindless Sensationalism (MS).

	Mindless Sensationalism is very similar in many
ways to the many-minds theories developed by Lockwood
\cite{Lo,Lo2,Lo96,Lockwood}
and by David Albert and Loewer
\cite{A,A92},
except that the basic conscious entities,
which Mindless Sensationalism asserts there are ``many'' of,
are conscious experiences rather than minds.

	By a ``conscious experience,'' I mean all that one is
consciously aware of or consciously experiencing at once.
Lockwood has called this a ``phenomenal perspective''
or ``maximal experience'' or ``conscious state.''
It could also be expressed as a total ``raw feel''
that one has at once.
In my papers on Sensible Quantum Mechanics
\cite{SQM,Page},
I have usually called it merely
a ``perception'' or sometimes an ``awareness''
or ``sensation,''
but I do not wish to imply that I
am using the same subtle meanings for those terms that others might.
For example, I am not merely considering
an individual sensory perception,
or even just the set of simultaneous sensory perceptions
of things external to the brain.
Instead, what I mean by a conscious experience or perception
is a total conscious awareness,
a ``subjective,'' ``internal,'' ``first-person'' experience
by roughly what one crudely thinks of as one conscious ``being,''
at roughly the one ``time'' that is then felt by the conscious ``being''
to be ``now.''
(However, I hasten to say that I
doubt the absolute existence
of any uniquely identifiable conscious ``beings'' within our universe,
and I also doubt the existence of
any entity with the precise properties commonly ascribed to ``time,''
except possibly for the admitted existence of mental concepts
within the contents of certain conscious experiences themselves.
For me the conscious experiences themselves
are the fundamental entities,
and it is only in trying to illustrate,
in commonly understood language, what I mean by them,
that I am apparently forced
to describe them in terms of what I regard as less
fundamental concepts such as ``conscious beings''
and ``at one time'' or ``now.'')

	A conscious experience can include components such as a
visual sensation, an auditory sensation, a pain, a conscious memory,
a conscious impression of a thought or belief, etc.,
that are all experienced together.  However, it does not
include a sequence of more than one immediate experience that in other
proposals might be considered to be strung together to form a stream of
consciousness of an individual mind.

	Because I regard the basic conscious entities to be
the conscious experiences themselves, which might
crudely be called sensations if one does not restrict
the meaning of this word to be the conscious responses
only to external stimuli, and because I doubt
that these conscious experiences are arranged in
any strictly defined sequences that one might define
to be minds if they did exist,
my framework has sensations without minds and hence
may be labeled Mindless Sensationalism.

	I should also emphasize that by a conscious experience,
I mean the phenomenal, first-person, ``internal'' subjective experience,
and not the unconscious ``external'' physical processes in the brain
that accompany these subjective phenomena.  In his first
chapter, Chalmers \cite{Chal} gives an excellent discussion of the
distinction between the former, which he calls the phenomenal concept
of mind, and the latter, which he calls the psychological concept of
mind. In his language, what I mean by a conscious experience
(and by other approximate synonyms that I might use,
such as perception or sensation or awareness)
is the phenomenal concept, and not the psychological one.

	Now that I have tried to illustrate
what I mean by the conscious experiences that
I take to be the basic entities that make up what might be called
the ``internal'' mental world
(which I shall here call the ``conscious world''),
let me turn to a quantum description of what
might be called the ``external'' physical world.
(This world I shall here call the ``quantum world''
to avoid offending the materialists who say
that consciousness is part of the ``material world,''
whatever that is supposed to mean, and to avoid offending
the physicists, myself included, who claim that
consciousness is part of the ``physical world,''
whatever that is supposed to mean---as a physicist
I shall take it to mean roughly
whatever is studied by those who consider themselves doing physics.
I'll nevertheless inevitably offend the smaller number of quantumists
who consider consciousness to be a quantum phenomenon,
but I want some short phrase to denote the non-conscious
aspects of a physics description of our universe,
without of course intending to deny that there is a relation
between consciousness and the quantum world.)

	For those who object that my terminology implies
an unrealistic dualism between the internal mental world
and the external physical world
(between the ``conscious world'' and the ``quantum world''
as I am using these terms),
I can say that I do not wish to imply that there
is necessarily a fundamental distinction between these two ``worlds,''
but at the present level of description it seems
to help to recognize the distinction between the two ways
of describing aspects of our universe.
Physicists often try to describe some aspects of our universe
by using the mathematical language of current physics
and ignoring consciousness,
and it seems that others (idealists?) can consider
conscious experiences separately from the aspects of our universe
that physicists usually consider.
There may be a deeper level of understanding
at which the ``conscious world'' and the ``quantum world'' are unified,
but to get to this level it does not seem to me to help
to pretend that at our present level of understanding
our descriptions do not usually make a distinction
between what appears to be these two different aspects of reality. 

	Rather than restricting attention to particular
theories or theoretical frameworks for the quantum world,
such as nonrelativistic quantum mechanics,
relativistic quantum field theory, quantum gravity,
or quantum string or M theory,
I shall here focus on what I consider to be
the basic elements of quantum theory as I presently understand it.

	In the Feynman path-integral approach,
the basic elements of quantum theory might
be a set of ``paths'' or fine-grained histories
allowed for the universe, and a rule for assigning
to each such history a complex number called an ``amplitude''
(a number of the form of a real number
plus $i$, the square root of $-1$, times another real number;
the complex number is itself real if the real number
multiplying $i$ is 0).
(The dynamical `theory of everything' or TOE would then
be primarily concerned with specifying the rule
for assigning the amplitudes, and the boundary conditions
or BC would essentially tell what paths are to be included.)
That is not the whole story in this approach, however,
as there seems to be a need to combine the individual paths
into appropriate sets of paths and add up the amplitudes
for all the paths in each set.
Precisely how this is to be done is a bit mysterious
to me, and so I find it a bit clearer to try
to relate consciousness more directly to another approach
to quantum theory, which might be called the operator approach.
(There are crude rules for which amplitudes
in the path-integral approach to add up
in practical situations, but I'm not sure these rules
are not implicitly invoking some assumptions about
something like consciousness, whereas at the level
of discussing only what I am calling the quantum world,
I would like to start with a set of structures that do not depend
in any way on consciousness.)

	In the operator approach,
the basic elements of quantum theory might
be a set of ``operators'' obeying some algebra
(rules for adding and multiplying them),
along with some ``quantum state'' (or simply ``state'')
for the universe
that determines a complex number called the
``expectation value'' for each operator.
I shall give some examples below, but for now
one can think of the operators as some abstract
mathematical entities that can be multiplied
by complex numbers, added or subtracted,
and multiplied together to give other operators.
The expectation value of the operators,
determined by the quantum state,
is required to be linear in that
the expectation value of the new operator
that is a certain complex number times the old operator
is simply that complex number times
the expectation value of the old operator,
and the expectation value of the sum of two operators
is simply the sum of the expectation values
of the two separate operators.
(However, the expectation value of the product
of two operators is not, in general,
the product of the expectation values
of the two separate operators.)

	In the operator approach, the operators
are somewhat analogous to the amplitudes for the paths
in the path-integral approach and so would be the part
primarily determined by the TOE.
Similarly, the quantum state is somewhat analogous to
the set of allowed paths in the path-integral approach
and so would be primarily determined by the BC.
Getting the expectation value for an operator would
be analogous to adding up the amplitudes for a certain
set of paths.  (Actually, on this issue the operator
approach seems a bit more complete,
since to say what an operator means in the path-integral
approach, one needs to say which set of paths
contribute to each operator,
usually with an additional complex weighting factor
besides the amplitudes for the paths themselves.)

	As an example of the operator approach,
consider the example of a `universe' consisting
of a single nonrelativistic particle moving in one
spatial dimension, e.g., along the $x$-axis.
In this simple case, the quantum states can
be represented by `wavefunctions' that are complex
functions of $x$, say $\psi(x)$, and which are
square-integrable, meaning in this case that
the integral of $|\psi(x)|^2$ over all $x$ is finite.
(The absolute value squared of a complex number,
such as $\psi = \psi_R + i\psi_I$ with
$\psi_R$ and $\psi_I$ being the two real numbers
that make it up, is $|\psi|^2 = \psi_R^2 + \psi_I^2$,
a real nonnegative quantity that is the square of
the distance, from the origin, of the point representing $\psi$
on the complex plane, which itself has a horizontal,
or `real,' axis representing the real part, $\psi_R$,
of the complex number $\psi$, and a vertical, or `imaginary,'
axis representing the imaginary part, $\psi_I$,
of the complex number $\psi$.)

	In this one-dimensional quantum example,
operators are mathematical entities that represent
ways of changing one wavefunction to another
in a linear way.  For example, corresponding
to the position $x$ that the particle might be considered
to have in a classical description,
there is the quantum position operator, say $X$, that
converts a wavefunction $\psi(x)$ to the wavefunction $x\psi(x)$.
(Strictly speaking, $X$ is not really a well-defined
operator if the space of states is represented by
all square-integrable wavefunctions, since
there exist square-integrable wavefunctions $\psi(x)$,
such as $\psi(x) = 1/\sqrt{\pi(1+x^2)}$,
for which $x\psi(x)$ is not square-integrable,
but to get a simple example, I shall here ignore the
mathematical technicalities that one can use
to get a class of wavefunctions for which $X$ is a good operator.)
Similarly, corresponding to the momentum $p$
that the particle might have in a classical description,
there is the quantum momentum, say $P$, that converts
a wavefunction $\psi(x)$ to the wavefunction $-id\psi(x)/dx$.

	Operators change states in linear ways,
so for complex numbers $a$ and $b$, the operator $aX+bP$
converts a wavefunction $\psi(x)$ to the wavefunction
$a x\psi(x) - ib d\psi(x)/dx$.

	The product of two operators, such as $PX$,
has the effect of performing the operations on the right
first, followed by the operation to the left.  Thus $PX$
converts a wavefunction $\psi(x)$ to the wavefunction
$-i d(x\psi(x))/dx = -ix d\psi(x)/dx - i \psi(x)$.
Note that, in general, the product of two operators
depends on the order in which the are taken, so $XP$
converts a wavefunction $\psi(x)$ to the wavefunction
$-ix d\psi(x)/dx$, the same as $PX - iI$ does,
where $I$ is the identity operator that
converts a wavefunction $\psi(x)$ to the same wavefunction $\psi(x)$.
This example shows that, for any wavefunction,
$XP = PX - iI$ or $PX-XP = iI$.
(This is a so-called commutation relation, since $PX-XP$,
which is mathematically denoted by $[P,X]$,
is called the commutator of $P$ and $X$.
This commutation relation is what essentially gives
the Heisenberg uncertainty relation for momentum and position,
but it would take me too far afield to explain that here.)

	Now that I have given an example of operators
from one-dimensional nonrelativistic quantum mechanics,
let me illustrate how quantum states give expectation values
to operators.  In Dirac's `bracket' notation, a `pure' quantum state
can written as the `ket' $|\psi\rangle$, which in my example
is represented by the wavefunction $\psi(x) = \psi_R + i\psi_I$,
or alternatively, it can be written as the `bra' $\langle\psi|$,
which is represented by the complex conjugate wavefunction
$\bar{\psi}(x) = \psi_R(x) - i\psi_I(x)$.
A slightly better representation of the pure state is the
combination $|\psi\rangle\langle\psi|$,
which avoids the phase ambiguity in representing
a pure state by either $|\psi\rangle$ or $\langle\psi|$
individually, since the state is physically the same
if $|\psi\rangle$ is multiplied by the complex phase factor
$e^{i\theta} = \cos{\theta} + i\sin{\theta}$
and $\langle\psi|$ is multiplied by the complex conjugate phase factor
$e^{-i\theta} = \cos{\theta} - i\sin{\theta}$
for some real angle $\theta$ measured in radians
(degrees divided by 180 and multiplied by $\pi$,
so that a 180-degree rotation is represented by
$\theta = \pi$, which gives $e^{i\theta} = e^{-i\theta} = -1$).
The phase factor has no physical consequences, and indeed
$|\psi\rangle\langle\psi|$ remains unchanged by it,
since $e^{i\theta}e^{-i\theta} = 1$ so that
$e^{i\theta}|\psi\rangle\langle\psi|e^{-i\theta}
=|\psi\rangle\langle\psi|$.
 
	The result of an operation, say $X$,
on a quantum state denoted by $|\psi\rangle$
can be then denoted as $X|\psi\rangle$, say $|\phi\rangle$,
and represented by the wavefunction $\phi(x)$
(which in this particular case is $x\psi(x)$).
Then the expectation value of $X$, denoted by $\langle X\rangle$,
is the `inner product' of the bra $\langle\psi|$
with the ket $|\phi\rangle = X|\psi\rangle$, which is
 \begin{equation}
 \langle X\rangle = \langle\psi|X|\psi\rangle
 = \langle\psi|\phi\rangle
 = \int_{-\infty}^{\infty}dx \bar{\psi}(x) \phi(x)
 = \int_{-\infty}^{\infty}dx \bar{\psi}(x) x \psi(x).
 \label{eq:1}
 \end{equation}
 
 	One can readily see from this example that the
expectation value is linear in the operators, e.g.
 \begin{equation}
 \langle aX+bP \rangle
  = a\langle X \rangle + b \langle P \rangle,
 \label{eq:2}
 \end{equation}
but in general,
 \begin{equation}
 \langle XP \rangle
 \neq \langle X \rangle \langle P \rangle.
 \label{eq:3}
 \end{equation}
 
	Although for my purposes below it is generally
sufficient to think of the universe as having a pure quantum state,
for completeness I should say that besides the pure states
best represented by the single term $|\psi\rangle\langle\psi|$,
one can have `mixed' or `statistical' states
represented by a sum of such terms,
 \begin{equation}
 \rho = \sum_{i,j}c_{ij}|\psi_j\rangle\langle\psi_i|,
 \label{eq:4}
 \end{equation}
with a set of different kets $|\psi_i\rangle$
and bras $\langle\psi_j|$,
where the $c_{ij}$'s form what is known as the density matrix,
which is Hermitian ($c_{ij} = \bar{c}_{ji}$),
positive (eigenvalues nonnegative), and
normalized (eigenvalues summing to unity).
For such a state, the expectation value of
an operator such as $X$ is
 \begin{equation}
 \langle X\rangle = tr(X\rho)
  = \sum_{i,j}c_{ij}\langle\psi_i|X|\psi_j\rangle.
 \label{eq:5}
 \end{equation}

	For infinitely large systems, there are
even more general states, known as C*-algebra states,
which need not be represented by normalized density matrices.
Instead, such states are represented by positive linear
functionals of the operators.
(A functional of a set of operators is something analogous to a formula
that gives a number for each operator.
A positive functional gives positive numbers for
positive operators, which are operators that have positive eigenvalues.
A linear functional gives a number for the sum of two operators
that is the sum of the two numbers that it would give
for each operator individually.)
If such a state is written symbolically as $\sigma$,
then one can write the expectation value of an operator
like $X$ as $\langle X\rangle = \sigma[X]$.
The pure and mixed states described above are
then special cases of these more general
C*-algebra states, so for generality we can denote any
quantum state by a positive linear functional $\sigma$.
 
 	So far I have not put time into the picture.
I believe that time is not a basic fundamental part
of physics, so in the ultimate description of the quantum world
(at least if it continues to use what I am here regarding as the
fundamental entities of quantum theory), there will be operators
and a quantum state for the universe, but no time.
However, in most of our approximate quantum theories for
models of parts of the universe, time does enter.
For example, in nonrelativistic quantum theory,
in the Heisenberg picture I shall use when speaking of time,
the quantum state is considered to be independent of time,
but the operators, like $X$ and $P$, are defined to be
functions of the time $t$, as $X(t)$ and $P(t)$.
(The wavefunction that represents the time-independent quantum state
$|\psi\rangle$ is then also a function of time, $\psi(x,t)$.)
Then the quantum algebra relates the operators at different times.
For example, for a free particle of unit mass,
the relation takes the simple form
 \begin{equation}
 X(t) = X(0) + t P(0)
 \label{eq:6}
 \end{equation}
and
 \begin{equation}
 P(t) = P(0).
 \label{eq:7}
 \end{equation}
The form of the relation of these operators at different times,
which I am considering to be part of the algebra of the operators,
depends on the dynamics of the system,
for example on the forces on the particle
in this simple one-dimensional example.
One might say that if time does not really exist,
then there is no dynamics,
which would trivialize the TOE,
but I take the attitude that it is the algebra
of the operators (the rules giving all their sums and products)
that represents the dynamics,
and this can persist even if time as we usually know it does not.

	I might add that even if one has time within some model
system, such as the one-dimensional nonrelativistic
quantum mechanical model described above,
if this system is really a closed quantum system,
what I believe is important about it is described
by the quantum state and the quantum operators,
but not the representation of the operators at various times.
For example, in Eq.~(\ref{eq:6}),
even though $X(t)$ has a different representation
from $X(0) + t P(0)$,
I believe there is fundamentally no distinction between them,
because they are equal operators.
Therefore, even in models in which a time such at $t$ exists,
the operators cannot be uniquely identified with any single time,
and so what I regard as the basic quantum entities
are effectively timeless.
Only if one augments the basic quantum theory of states
and operators with distinctions between different forms
of the same operators, such as the left and right hand sides
of Eq.~(\ref{eq:6}), does one get any real dependence upon
the time parameter $t$.

	If the quantum world is described by operators
and states (with our universe being described by
one particular set of operators and by a particular state,
the so-called `quantum state of the universe'),
then a goal of a theory of psycho-physical parallelism (PPP)
would be to give the connection between
the quantum state of the universe
and the conscious experiences occurring within it.
Eschewing the extreme solipsistic view that only
my present conscious experience exists,
I assume that many conscious experiences exist within the universe,
so a PPP should give many conscious experiences
for a single quantum state.

	Suppose that one denotes an individual conscious experience
by the letter $p$ and defines the ``conscious world'' $M$
as the set of all possible conscious experiences in all universes
with all possible quantum states (i.e., not just in our universe
with its particular quantum state $\sigma$).
One logically possible view would be that all possible conscious
experiences exist equally, regardless of the quantum state.
But this would make the quantum world completely irrelevant
for the existing conscious experiences,
and so the apparent order that I sense within my present
experience would not at all be explained by any postulated
quantum world.
On the contrary, I feel that the order that I sense
within my own experience is better explained
by assuming that there is a quantum world and that
the conscious experiences are in some sense correlated with it.
Therefore, I shall make this assumption,
that there is indeed a nontrivial psycho-physical parallelism.

	The next possibility one can consider is
the assumption that the quantum state of the universe
restricts the set of conscious experiences that actually exist
to be a proper subset, say $E$, of the set $M$
of all possible conscious experiences,
but that each conscious experience within the existing set $E$
is equally real.
This would seem to be a reasonable assumption if
the quantum world were actually classical,
so that some physical possibilities definitely happen and others do not.
Then it would be plausible that some conscious experiences
definitely happen and that others do not.

	For example, suppose that one takes a simplified
nonrelativistic classical model in which there are a certain set
of pointlike elementary particles that move along definite
trajectories through space as a function of time,
so that at each time there is a definite configuration
of the positions of these particles in space.
The temporal sequence of these configurations
could then be called the classical history of this universe. 
Certain sets of the configurations might be identified
as conscious brain states, and for each of these
one might identify a corresponding conscious experience $p$.
Then one might propose that if a configuration corresponding
to the conscious experience $p$ occurs during the classical history
of this universe, then this conscious experience exists,
but if the configuration never occurs, the corresponding
conscious experience does not exist either.
If there is some correspondence between the orderliness of
the physical brain configurations and the orderliness
perceived within the conscious experience,
then an orderly history could explain orderly conscious experiences.

	A similar picture with a definite sequence of configurations
occurs in the deBroglie-Bohm version of nonrelativistic quantum theory
\cite{deB,Bohm,Hol,BDDGZ,CFG},
in which to the normal operators
and state there is added a definite trajectory
whose evolution, but not whose initial configuration,
is determined by the wavefunction, which acts as a `pilot wave.'
However, it seems to me unnecessary to add a trajectory
to quantum theory, which for completeness would require
a specification of its initial configuration as additional information.
It also seems very ugly to try to do this for examples
beyond nonrelativistic quantum mechanics.
For example, in relativistic quantum field theory,
a trajectory of sequences of field configurations
that obeyed Bohm's equation for the evolution of the configurations
using the time corresponding to one observer would not obey
that equation using instead the time corresponding to a moving observer,
so that relativistic invariance would be broken by the trajectories.

	However, in a quantum theory with operators and a state,
unless one adds extra elements like the definite trajectory
of Bohm's version of quantum theory,
it seems difficult or ugly to have the operators and state
give a definite rule for saying that some possible conscious
experiences definitely exist but that others do not.
It is much easier to have a rule assigning different (nonnegative real)
weights or levels of reality to different conscious experiences,
with the rule depending upon the quantum state of the universe.
Then if all conscious experiences with positive weights $w$
are said to exist, but if experiences with greater weights
exist in some sense more, then one might expect that it is more
likely that one's experience would be one that has greater weight.
(One might like to propose that one simply takes all possible conscious
experiences with positive weight as existing and all possible conscious
experiences with zero weight as not existing, but for the simplest
ways of assigning the weights from quantum theory,
such as what I shall give below,
almost all of the possible conscious experiences would have
a weight at least a tiny bit positive, so this proposal would
exclude as nonexisting only an infinitesimally small fraction
of the total set $M$ of conscious experiences $p$.
Therefore, I am not considering this particular proposal further.)

	In other words, if the weight $w(p)$ gives the level of reality
or existence of the conscious experience $p$,
one can say that in the universe almost all possible conscious
experiences exist in the sense of having
at least some positive measure of reality,
but some sets of experiences are much more real than others,
existing to a much greater degree than other sets.
One way to describe this is to imagine randomly selecting
a conscious experience $p$ out of all of the possible ones.
For a random selection one always needs a weight,
and if it is chosen to be the weight $w(p)$ that comes
from the quantum state $\sigma$ by some particular
theory of psycho-physical parallelism,
then the probability that a particular conscious experience $p$
will be chosen by the random selection will be proportional to
its weight $w(p)$.

	In this way one can say that the weight $w(p)$
is analogous to the probability for the conscious experience $p$,
but it is not to be interpreted as the probability
for the bare {\it existence} of $p$, since any conscious experience
$p$ exists (is actually experienced) if its weight is positive,
$w(p) > 0$.  Rather, $w(p)$ is to be interpreted as being
proportional to the probability of {\it getting} this particular
experience if a random selection is made.

	A more picturesque way of viewing the weight,
but one which has the danger of misinterpretation if
all of the elements used in the picture are assumed to have reality
or are confused with similar elements that occur
in our present approximate theories of the world,
is the following analogy:
Assume that God has His own time (not to be confused with
the time that we use in our present approximate physical theories,
but having some properties analogous to what we often
assume, perhaps erroneously, that time does in our
approximate physical theories),
and that as He creates each conscious experience,
He spends a time $w(p)$ giving existence to each.
In other words, assume that each exists for an amount $w(p)$
of God's time.
Then the conscious experiences with greater $w(p)$
will have a greater existence in the sense of their duration
in God's time.
The picture is then that the weight
for conscious experiences may be viewed as
somewhat analogous to the measure of physical time
used for calculating time averages in dynamical systems, for example.

	Because the specification of the conscious experience $p$
completely determines its content and how it is experienced
(how it feels), the weight $w(p)$ has absolutely no effect
on that---there is absolutely no way within the experience
to sense anything directly of what the weight is.
A toothache within a particular conscious experience $p$
is precisely as painful an experience no matter what $w(p)$ is.
Furthermore, the experience $p$ is whatever $p$ is
and has absolutely no memory of how long God may have
had that experience existing within His time in the analogy.
It is just that an experience with a greater $w(p)$
is more likely in the sense of being more probably
chosen by a random selection using the weights $w(p)$.
(Of course, the experience $p$ might include a conscious
awareness of belief in a theory that assigns a particular weight
to that experience, but the awareness of that belief will be
part of $p$ itself and will not directly depend
on whether the actual weight is what the believed theory assigns
for it.  In this way a conscious belief depends only on the
conscious experience of which that belief is a part
and not on the truth of the implications of that belief.
It is only by faith in the orderliness of the universe
that we can assume that our conscious orderly beliefs
about it are true, and even that faith itself
can be regarded to be just given as part
of the corresponding conscious experience.)

	If one takes the attitude that there is no reality
to a divine temporal period $w(p)$ for the existence of
the conscious experience $p$ (in the analogy that admittedly
is rather contrived), and that there is no reality
to the random selection with weights proportional to $w(p)$,
then one might think that the weights have no reality
but are merely a meaningless arbitrary assignment.
I do find it difficult to try to describe the weights
in terms of anything more basic
and of whose existence I am more confident,
but I also believe that the weights really are fundamental
elements of reality.  In other words, I believe that some
sets of conscious experiences
really do have a greater measure of reality
than others, and this greater measure is the explanation
of why my present experience has its experienced orderliness:
such orderly experiences have greater weight than ones
which are much more disorderly.
Of course, I cannot prove this assumption,
but it enables me to make progress toward finding
an explanation of the orderliness that I experience,
so I shall continue to make it here.

	One technical point that it is now time to make
is that to simplify the discussion above, I have often
implicitly assumed that the set $M$ of possible conscious
experiences is a countable discrete set,
so that, for example, one can imagine choosing
an experience $p$ at random with weight $w(p)$.
In particular, if the total sum of the weights
for all conscious experiences is finite and is normalized to be unity,
then the weight $w(p)$ for each conscious experience is simply
the probability for that experience to be chosen
by the random selection.
This is indeed the possibility that is the easiest to
visualize, and it generally will not hurt to have it in mind
for most of the discussion below,
but in forming a fairly general framework for the connection
between the quantum and mental worlds,
I would not like to make unnecesary restrictions,
and so I shall allow the possibility that the set of
conscious experiences may be uncountable or continuous.
(Is there a true continuum for the pain of a toothache,
or are there only a countable set of discrete values
for how painful it can be experienced?
We don't know which it is, so I shall allow either possibility.)

	If the set $M$ of conscious experiences
is a continuum, then a nonzero weight for a single conscious experience
$p$ (a point in this continuum) is rather meaningless,
but in reasonable cases one can still have a weight
for any set $S$ of experiences, even if this weight is
zero for any single individual experience.
(For even this to be possible, the set $M$ of all
possible experiences must be a measurable set,
which I shall continue to assume,
since I personally don't know how to make much sense
of a generalization in which that is not true.)
To give the weight for a set of experiences a fancier name,
let us henceforth call it the {\it measure}
$\mu(S)$ of the subset $S$ of the full set $M$ of possible
conscious experiences.

	Then one can imagine that if exclusive subsets
are being selected randomly with the measure $\mu$,
then the ratio of the probability of choosing $S_1$, say,
to that of choosing $S_2$ would be $\mu(S_1)/\mu(S_2)$,
so the measures for the sets would give their relative probabilities.
If $\mu(M)$ is finite, then one can define a normalized
weight $P(S)=\mu(S)/\mu(M)$ which would be the probability
of choosing the subset $S$ if one randomly selected, with the
measure $\mu$, among an exhaustive and exclusive set of subsets
of $M$ that includes the subset $S$.
For example, if $S_1$ is the set of conscious experiences in which
no toothache is felt and $S_2$ is the set of conscious experiences
in which a toothache is felt, then these two subsets of $M$
form an exhaustive and exclusive set of two subsets of $M$,
since every conscious experience $p$ in $M$ is in $S_1$
or $S_2$ (exhaustive subsets), and no experience is in both
(exclusive subsets).  Therefore, $\mu(M) = \mu(S_1) + \mu(S_2)$,
and $P(S_2) = \mu(S_2)/\mu(M)$ is the probability
of randomly selecting a conscious experience with a toothache.

	However, it might be that the total set $M$
of conscious experiences is so large,
and the measure $\mu(S)$ for its subsets $S$
is so widely spread, that the total measure of $M$
is divergent.
(A simple example would be if $M$ could be put into
one-to-one correspondence with the real number line,
$-\infty < x < \infty$, and if the measure for the set
$S = \{x|x_1 < x < x_2\}$ were $\mu(S) = x_2 - x_1$,
simply the length of the interval for $x$.)
Then any subset with finite measure $\mu(S)$ would have zero
absolute probability of being chosen if one divided
by the infinite $\mu(M)$.
Also, even if one chose subsets with infinite
measure, dividing that infinity by the infinity
of the total measure would generally give ambiguous results,
and so absolute probabilities that are not zero would be
ambiguous.  This might make it hard to test such a theory.
However, if one had two subsets with finite measure,
say $S_1$ and $S_2$, then one would get a finite conditional
probability to be in, say, $S_1$, given that one is in
the union of the two sets, and so there still might be some
tests of such a measure that one could make.
Therefore, I am hesitant at this stage to demand
that the total measure for the full set $M$
of all possible conscious experiences be finite.

	Now, having explained briefly what I take the basics
of quantum theory to be and what it might mean to have
a set of conscious experiences with a measure,
it is time to write these as axioms and add my axiom
for the basic structure of the psycho-physical parallelism.

	Mindless Sensationalism (MS) is given by the
following three basic postulates or axioms
\cite{SQM}:

 {\bf Quantum World Axiom}:  The unconscious ``quantum world'' $Q$ is
completely described by an appropriate algebra of operators and by a
suitable state $\sigma$ (a positive linear functional of the operators)
giving the expectation value $\langle O \rangle \equiv \sigma[O]$ of
each operator $O$.

 {\bf Conscious World Axiom}:  The ``conscious world'' $M$,
the set of all conscious experiences or perceptions $p$,
has a fundamental measure $\mu(S)$ for each subset $S$ of $M$.

 {\bf Psycho-Physical Parallelism Axiom}:
The measure $\mu(S)$ for each
set $S$ of conscious experiences is given by the expectation value of a
corresponding ``awareness operator'' $A(S)$, a positive-operator-valued
(POV) measure, in the state $\sigma$ of the quantum world:
 \begin{equation}
 \mu(S) = \langle A(S) \rangle \equiv \sigma[A(S)].
 \label{eq:8}
 \end{equation}

	For $A(S)$ to be a POV measure, it is necessary that
$A(S)$ be zero when $S$ is the empty set and otherwise
be either zero or else a positive operator,
which implies that $\sigma[A(S)] \geq 0$ for all
positive linear functionals $\sigma$,
and it is also necessary that if the set $S$ is a countable union
of disjoint sets $s_i$, $A(S)$ is the sum of the $A(s_i)$
when this sum ``converges in the weak operator topology''
\cite{Dav}.  Then $\mu(S)$ has the standard
additivity property of a measure.

 	As essentially mentioned above in my description
of what I consider to be the basics of quantum theory,
the Quantum World Axiom is here deliberately vague as to the precise
nature of the algebra of operators and of the state, because as the
details of various quantum theories of the universe are being
developed, I do not want the general framework of Sensible Quantum
Mechanics at this time to be made too restrictive.

	The Psycho-Physical Parallelism Axiom states my assumption
of the structure of the `psycho-physical laws,'
the laws that presumably give the `neural correlates of consciousness.'
This axiom, when combined with the other two, gives what to me seems
to be the simplest and most conservative framework for ``{\it bridging}
principles that link the physical facts with consciousness'' and for
stating ``the connection at the level of `Brain state X produces
conscious state Y' for a vast collection of complex physical states and
associated experiences'' \cite{Chal} in language that is consistent
with Sydney Coleman's description
\cite{Sid93,Sid95}
of quantum theory as having ``NO special measurement process, NO
reduction of the wavefunction, NO indeterminacy''
(in particular,
with a many-experiences variant of Everett's quantum theory
\cite{E,DG},
in which measures for sets of conscious experiences are added
to the bare unitary quantum theory that Coleman advocates).

	The Psycho-Physical Parallelism Axiom is the simplest way
I know of connecting the quantum world with the conscious world.
One could easily imagine more complicated connections,
such as having $\mu(S)$ be a sum or integral,
over the conscious experiences $p$ in the set $S$,
of some nonlinear function of the expectation values, say $m(p)$,
of positive ``experience operators'' $E(p)$ depending in the $p$'s
\cite{SQM}.
Instead, my Psycho-Physical Parallelism Axiom
restricts the functions in the sum or integral
to be linear in the expectation values.
In short, I am proposing that the psycho-physical parallelism
is {\it linear}.

	Of course, the Psycho-Physical Parallelism Axiom, like the
Quantum World Axiom, is here also deliberately vague
as to the form of the awareness operators $A(S)$, because I do not
have a detailed theory of consciousness, but only a framework for
fitting it with quantum theory. My suggestion is that a theory of
consciousness that is not inconsistent with bare quantum theory should
be formulated within this framework (unless a better framework can be
found, of course). I am also suspicious of any present detailed theory
that purports to say precisely under what conditions in the quantum
world consciousness occurs, since it seems that we simply don't
know yet. I feel that present detailed theories may be analogous to the
cargo cults of the South Pacific after World War II, in which an
incorrect theory was adopted, that aircraft with goods would land
simply if airfields and towers were built.

	Since all sets $S$ of conscious experiences
with $\mu(S) > 0$ really occur
in the framework of Mindless Sensationalism, it is completely
deterministic if the quantum state and the $A(S)$ are determined: 
there are no random or truly probabilistic elements in MS.
Nevertheless, because the framework has measures for sets of
conscious experiences, one can readily use them to calculate quantities
that can be interpreted as conditional probabilities.
One can consider sets of conscious experiences $S_1$, $S_2$, etc.,
defined in terms of properties of the conscious experiences.
For example, $S_1$ might be the set of conscious experiences in
which there is a conscious memory of having tossed a coin one hundred
times, and $S_2$ might be the set of conscious experiences
in which there is a conscious memory of getting more than seventy heads.
Then one can interpret
 \begin{equation}
 P(S_2|S_1)\equiv \mu(S_1\cap S_2)/\mu(S_1)
 \label{eq:9}
 \end{equation}
as the conditional probability that the
conscious experience is in the set $S_2$,
given that it is in the set $S_1$.
In our example, this would be the conditional probability that a
conscious experience included a conscious memory
of getting more than seventy heads,
given that it included a conscious memory of having tossed a
coin one hundred times.

	An analogue of this conditional ``probability''
is the conditional probability that a person
at the beginning of the 21st century is the Queen of England.
If we consider a model of all the six billion people,
including the Queen, that we agree to consider as living humans
on Earth at the beginning of 2001,
then at the basic level of this model the Queen certainly exists in it;
there is nothing random or probabilistic about her {\it existence}.
But if the model weights each of the six billion people equally,
then one can in a manner of speaking say that
the conditional probability that one of these persons is the Queen
is somewhat less than $2\!\times\! 10^{-10}$.
I.e., if one chooses at random one of the six billion people
on Earth at the beginning of 2001, with each person
being assigned an equal probability of being chosen,
then the probability of {\it getting} the Queen
by this random selection is, to one-digit accuracy,
$2\!\times\! 10^{-10}$.
(One can see that this probability of getting the Queen
would be much more if one instead weighted
the probability for each person by the weight of his or her crown,
which would be analogous to having a different
quantum state giving a different $\mu(S) = \langle A(S)\rangle$.)
I am proposing that it is in the same manner of speaking
that one can assign conditional probabilities
to sets of conscious experiences,
even though there is nothing truly random about them at the basic level.

	As it is defined by the three basic axioms above,
Mindless Sensationalism is a framework
and not a complete theory for the universe,
since it would need to be completed by giving the detailed algebra of
operators and state of the quantum world, the set of all possible
conscious experiences of the conscious world,
and the awareness operators $A(S)$
for the subsets of possible conscious experiences,
whose quantum expectation values are the measures for these subsets.

	Furthermore, even if such a complete theory were found,
it would not necessarily be the final theory of the universe,
since one would like to systematize the connection
between the elements given above.
As Chalmers eloquently puts it on pages 214-15 of his book \cite{Chal},
``An ultimate theory will not leave the connection at the level of
`Brain state X produces conscious state Y' for a vast collection of
complex physical states and associated experiences. Instead, it will
systematize this connection via an underlying explanatory framework,
specifying simple underlying laws in virtue of which the connection
holds.  Physics does not content itself with being a mere mass of
observations about the positions, velocities, and charges of various
objects at various times; it systematizes these observations and shows
how they are consequences of underlying laws, where the underlying laws
are as simple and as powerful as possible. The same should hold of a
theory of consciousness. We should seek to explain the supervenience of
consciousness upon the physical in terms of the simplest possible set
of laws.

	``Ultimately, we will wish for a set of {\it fundamental laws}.
Physicists seek a set of basic laws simple enough that one might write
them on the front of a T-shirt; in a theory of consciousness, we should
expect the same thing.  In both cases, we are questing for the basic
structure of the universe, and we have good reason to believe that the
basic structure has a remarkable simplicity. The discovery of
fundamental laws may be a distant goal, however. \ldots

	``When we finally have fundamental theories of physics and
consciousness in hand, we may have what truly counts as a theory of
everything. The fundamental physical laws will explain the character of
physical processes; the psychophysical laws will explain the conscious
experiences that are associated; and everything else will be a
consequence.''

	Returning to the elements above of a postulated completed,
but not necessarily final, Mindless Sensationalism theory,
it is presently premature to try to give these elements precisely,
particularly the awareness operators that have generally been left out
of physics discussions.  However, one might give a crude discussion
of what they might be like in some highly approximate way.

	One very strong assumption that might possibly be plausible
for certain quantum theories, is what I have called
the Orthogonal Projection Hypothesis
\cite{SQM}.
In the terms of the present paper,
this implies that the awareness operators $A(S)$
are projection operators, say $\Pi(S)$
(operators which remain the same when multiplied by themselves:
$\Pi\Pi=\Pi$,
which implies that the eigenvalues of the operator
are either zero or one),
and that the awareness operators for two disjoint
sets of conscious experiences, say $S_1$ and $S_2$,
are orthogonal, so $A(S_1)A(S_2) = A(S_2)A(S_1) = 0$.
(I should say that I see several reasons for doubting
that this very strong Commuting Projection Hypothesis
is really plausible
as a precise condition on the awareness operators,
so I am not advocating this assumption as the final word,
but it might be approximately true at least for
certain sets $S$ of conscious experiences,
and it does lead to various simple consequences.)

	A projection operator corresponds to
a corresponding property that a state may have with certainty
(if it is an eigenstate of that operator with unit eigenvalue)
or that a state may be certain not to have
(if it is an eigenstate of that operator with zero eigenvalue).
For a given projection operator, a generic state
is not an eigenstate and so is not considered with certainty
either to have the property or not to have it.
This is an expression of what is often considered the
uncertainty of quantum theory, though I would just regard it
as a limitation on what ``certain'' properties a system has.

	In the Copenhagen version of quantum theory,
to which I do not subscribe except
in a very rough instrumentalist sense,
a `measurement' is assumed to cause a normalized quantum state
to change or `collapse' to another quantum state given by
applying a projection operator to the original state
and then renormalizing its magnitude.
The expectation value of the projection operator,
$P = \langle \Pi \rangle$
in the original state, is then interpreted
as the probability that that state will thus collapse,
effectively giving a ``yes'' answer to the question
posed by the measurement of whether
the system being measured has the property corresponding
to the projection operator $\Pi$.
($1-P = \langle (I-\Pi) \rangle$
is then the probability that the answer will be ``no,''
so that the state will instead
collapse to the other possibility,
which is that given by applying the
complementary projection operator $I-\Pi$
to the original state and renormalizing it---here
$I$ is the identity operator that leaves a state the same.)
The fact that $\Pi$ is a projection operator
means that if the state collapsed to the ``yes'' answer,
a second measurement of precisely the same property
would with certainty give the answer ``yes'' again,
so that after the state collapses the first time, to
an eigenstate of the projection operator with unit eigenvalue,
the property corresponding to the projection operator will
with certainty be true.

	To illustrate projection operators, return to the example
of a single nonrelativistic particle moving along the $x$ axis,
with its quantum state represented by a wavefunction $\psi(x)$
which is normalized so that
the integral of $|\psi(x)|^2$ over all $x$ is unity.
In this case a simple example of a projection operator $\Pi$
is one which determines whether the particle is in some range
of $x$, say the range $x>0$.  The expectation value of this is then
$P$, the integral of $|\psi(x)|^2$ over all positive $x$,
and if the quantum state collapses to this possibility
in the Copenhagen version of quantum theory,
the wavefunction would change to $\psi(x)/\sqrt{P}$
for $x>0$ and to $0$ for $x<0$,
effectively giving a ``yes'' answer to the measurement
determination of whether the particle was to the right of the origin.
On the other hand, if the answer is ``no,'' which would occur
with a probability $1-P$, the wavefunction would change to
$0$ for $x>0$ and to $\psi(x)/\sqrt{1-P}$ for $x<0$.
This change is known as the `collapse of the wavefunction'
or the `reduction of the quantum state.'

	In my Mindless Sensationalism,
the quantum state of the universe never changes
by any collapse or reduction mechanism.
However, if the awareness operator $A(S)$
for a certain set of conscious experiences is
a projection operator $\Pi$,
and if the quantum state is normalized so that
the expectation value of the unit operator $I$ is unity, then
$\mu(S) = \langle A(S) \rangle = \langle \Pi \rangle = P$,
the same as the probability in the Copenhagen version of quantum theory
that measuring the property corresponding to
$\Pi$ would give a ``yes'' answer.

	For example, it is tempting to suppose that if
the set of conscious experiences is a set of very similar
experiences (or perhaps just a single experience
if the set of possible experiences is countably discrete)
that would occur for a person having a particular
brain configuration, then $A(S)$ is approximately
a projection operator onto those brain configurations.
In this case, the measure $\mu(S)$ for those experiences
would then be the same as the probability for the corresponding
brain configurations in Copenhagen quantum theory.

	The Orthogonal Projection Hypothesis appears to be a specific
mathematical realization of part of Lockwood's proposal
\cite{Lo} (p. 215),
that ``a phenomenal perspective [what I have here been usually calling
a conscious experience $p$] may be equated with a shared eigenstate of
some preferred (by consciousness) set of compatible brain
observables.''  Here I have expressed the ``equating'' by
my Quantum-Consciousness Connection Axiom,
and presumably the ``shared eigenstate''
can be expressed by a corresponding projection operator $\Pi$.

	Or, as Lockwood has expressed it in this present volume
\cite{Lockwood},
``I am suggesting, in other words, that the contents of consciousness,
at any given moment, correspond to a set of measurement outcomes
that belong to the respective spectra of a compatible set of observables
on the mind, construed as a subsystem of the brain.''
If this suggestion is incorporated within my axioms,
it effectively assumes that the awareness operators
corresponding to sets of conscious experiences
``at any given moment'' obey the Orthogonal Projection Hypothesis.
However, in my axioms I do not need a definition of what
``at any given moment'' might mean,
and I do not need to be able to define the mind as a subsystem
of the brain; for me the awareness operators $A(S)$ are basic.
(I also do not need the Orthogonal Projection Hypothesis,
though for now it is interesting to examine the consequences
if it were true.)

	I should also emphasize that if the same conscious experience
is produced by several different orthogonal ``eigenstates of
consciousness'' (e.g., different states of a brain and surroundings
that give rise to the same conscious experience $p$), then in the
Orthogonal Projection Hypothesis the projection operator
$\Pi$ would be a sum of the
corresponding rank-one projection operators and so would be a
projection operator of rank higher than unity.  This is what I would
expect, since surely the surroundings could be different and yet the
appropriate part of the brain, if unchanged, would lead to the same
experience.
As Lockwood has put it
\cite{Lockwood},
``In particular, the contents of consciousness would seem to be
highly {\it coarse-grained}, in relation to the immensely
intricate physical processes on which they ostensibly supervene.
This difficulty for materialism was taken very seriously by the
philosopher Wilfred Sellars
\cite{Sel},
who dubbed it the `grain problem'. \ldots\
Crucially, I also assume that the compatible set of observables,
corresponding eigenvalues of which jointly define a given state
of consciousness, is a less than {\it complete} set.
A complete compatible set of observables is one that,
when measured, yields {\it maximal information}
concerning the measured system---information that cannot
be improved on by adding further observables to the set.
This relates directly to the grain problem.  Only by allowing the
operative compatible sets of observables to be {\it incomplete}
can we ratchet down the degree of resolution and complexity
of the corresponding conscious state to what one would
intuitively judge to be the right level.''

	On the other hand, if $A(S)$ were a sum of noncommuting
projection operators, or even a sum of commuting projection operators
that are not orthogonal, or if it were a
weighted sum of orthogonal projection operators
with weights different from unity,
then generically $A(S)$ would not be a projection operator
$\Pi$ as assumed in the Orthogonal Projection Hypothesis.
Although it would mean that the situation would not be so simple
as one (e.g., Lockwood, or I in an optimistic moment)
might like to assume, I see no fundamental difficulty
in having the awareness operators not be projection operators
and not be orthogonal.

	There are many other alternative technical assumptions that one
might make about the awareness operators
\cite{SQM},
but I shall not discuss them further here.

	Another point I should emphasize is that in
Mindless Sensationalism, there is no fundamental
notion of a correlation between distinct conscious experiences.
One can get the measure (and the normalized probability,
if the total measure for the set $M$ of all conscious experiences
is finite) for any set $S$ of experiences,
but one does not get any nontrivial fundamental formula for
the joint occurrence of distinct experiences.
In particular, there does not seem to be any fundamental formula
for the conditional probability of one set $S$ of experiences
given a second set $S'$ that is exclusively distinct,
having no elements in common with the first set $S$
(other than the formula for the basic probability $P(S)$
of the first set,
the trivial conditional probability).
This essentially fits the crudely-expressed fact
that by the definition of a conscious experience $p$,
a ``conscious being'' can be directly aware of only ``one at a time.''
From the memory components of a ``present'' experience,
one might postulate the existence of a ``past'' experience
in which what is now just remembered is at that ``past'' ``time'' then
experienced as occurring simultaneously with the ``past'' experience
itself when that experience was being experienced.
However, within one's present experience,
one has no direct experience of the past experience itself.
Correspondingly, within my framework of Mindless Sensationalism,
there is no fundamental way to assign a probability
of a ``past'' experience given a particular present one.
Instead, each experience (if countably discrete, or else
each set of experiences if one must combine a continuum of them
to get a nonzero measure $\mu(S)$) has its own measure,
which is independent of the realization of any other experiences.

	In the other direction of ``time,''
Mindless Sensationalism does not assign any fundamental
conditional probabilities to any ``future'' experiences
given the existence of a particular present one.
One might think that it should,
since it is just common sense that probabilities for
the future depend upon present conditions.
For example, in Copenhagen quantum theory,
if the quantum state of the universe collapses to, say,
an eigenstate with unit eigenvalue of one of a particular
$A(S)$ that is, say, a projection operator $\Pi$,
then one expects that the probabilities of future conscious
experiences will depend upon which $A(S)$ the quantum state
collapsed to.
If the quantum state collapses to an eigenstate of the
assumed projection operator in which you are aware of winning
a large lottery,
one would expect that a month later, the probability
that you would experience an awareness of having a lot of money
would be greater than if the quantum state collapsed to
an eigenstate in which you were not aware of winning any large lottery
(assuming that you would not spend most of the money within the month).

	However, in Mindless Sensationalism,
the measure or probability of any ``future'' conscious experience
is completely determined by the (full) theory and is independent
of the occurrence of any ``present'' conscious experience.
This sounds absurd.  How can it be reconciled with our experience?
I am aware of having a computer in front of me;
isn't this correlated with my past awareness of buying a computer?

	The answer is that this experience does not show
any correlations between different experiences
(e.g., between those at different ``times'')
but rather the correlations between the different components
of a single present experience
(e.g., of perceiving a visual image of a computer screen
and of being consciously aware of a memory of buying the computer).
These are the correlations to be explained by a full theory
of Mindless Sensationalism.  (I'm just giving the framework here;
the full theory will involve an enormous amount of work,
and I suspect that humans will never completely develop it,
though I hope they will learn a lot more about it
than the pittance we know now,
and perhaps even develop an approximate outline of it.)

	Similarly, a prediction of what might seem to be
a correlation between a ``present'' awareness of winning a large lottery
and a ``future'' awareness of having a lot of money is,
I would claim, not that at all, but rather a prediction
of a correlation between one's ``future'' awareness of
having a lot of money and, within the same conscious experience,
a conscious awareness of a memory of having won a large lottery.

	To give another example, I can predict that if you are
consciously aware of reading this paper today
(i.e., if you are not reading it in a daze,
with no conscious awareness of what you are doing,
though I am not claiming that reading it unconsciously is impossible
or even that this possible experience is uncorrelated
with the content of this paper),
you will consciously remember my phrase
``Mindless Sensationalism'' tomorrow
if you think about my paper then.
Am I predicting something about your experience tomorrow
that is conditional upon your experience today?
No.  I am just predicting that in your conscious experience
of remembering reading this prediction of mine the day before,
within the same conscious experience there will be a reasonably high
probability that you will also be aware of my phrase
``Mindless Sensationalism.''

	The fundamental timelessness of Mindless Sensationalism
seems to fit very well with the viewpoint eloquently expressed
by Julian Barbour
\cite{Bar},
that ``Heraclitan flux \ldots may well be nothing but
a well-founded illusion.''
(I might note that although I almost entirely agree with
what Barbour writes, I am perhaps not quite
such as extreme anti-temporalist in that I suspect
that the quantum state of the universe may be given
by a path integral that has something analogous
to histories in them,
even though I agree with Barbour that the universe fundamentally
does not have anything like a classical history or classical time.
I think Barbour would also agree with me that
there are no fundamental sequences of conscious experiences.)

	In saying that Mindless Sensationalism
posits no fundamental correlation
between complete conscious experiences,
I do not mean that it is impossible to define such
correlations from the mathematics, but only that I do not see any
fundamental physical meaning for such mathematically-defined
correlations.  As an example of how such a correlation might be
defined, consider that if an awareness operator $A(S)$ is a projection
operator, and the quantum state of the universe is represented by the
pure state $|\psi\rangle$,
one can ascribe to the set of conscious experiences
$S$ the pure Everett ``relative state''
\cite{E,DG}
 \begin{equation}
 |S\rangle=\frac{A(S)|\psi\rangle}{\parallel 
 A(S)|\psi\rangle\parallel}
 =\frac{A(S)|\psi\rangle}
 {\langle\psi|A(S)A(S)|\psi\rangle^{1/2}}.
 \label{eq:10}
 \end{equation}
Alternatively, if the quantum state of the universe is represented
by the density matrix $\rho$, one can associate the set of experiences
$S$ with a relative density matrix
 \begin{equation}
 \rho_S=\frac{A(S)\rho A(S)}{Tr[A(S)\rho A(S)]}.
 \label{eq:11}
 \end{equation}
Either of these formulas can be applied when the awareness operator
$A(S)$ is not a projection operator,
but then the meaning is not necessarily so clear.

	Then if one is willing to say that $Tr[A(S)\rho]$ is the
absolute probability for the set of experiences $S$
(which might seem natural at least when $A(S)$ is a projection operator,
though I am certainly not advocating this na\"{\i}ve interpretation,
and in general it will not agree in absolute magnitude
with $P(S) = \mu(S)/\mu(M)$),
one might also na\"{\i}vely interpret $Tr[A(S')\rho_S]$
as the conditional probability of the set of experiences $S'$
given the set of experiences $S$.

	Another thing one can do with two sets of experience
$S$ and $S'$ is to calculate an ``overlap fraction'' between them as
 \begin{equation}
 f(S,S')=\frac{\langle A(S)A(S')\rangle\langle  A(S')A(S)\rangle}
 {\langle A(S)A(S)\rangle\langle A(S')A(S')\rangle}.
 \label{eq:12}
 \end{equation}
If the quantum state of the universe is pure, this is the same as the
overlap probability between the two Everett relative states
corresponding to the two sets of experiences:
$f(S,S')=|\langle S|S'\rangle|^2$. 
Thus one might in some sense say that if $f(S,S')$ is near unity, the
two sets of experiences are in nearly the same one of the Everett
``many worlds,'' but if $f(S,S')$ is near zero,
the two conscious experiences are in nearly orthogonal different worlds.
However, this is just a manner of speaking,
since I do not wish to say that the quantum state of the
universe is really divided up into many different worlds.
In a slightly different way of putting it,
one might also propose that $f(S,S')$, instead of $Tr[A(S')\rho_S]$,
be interpreted as the conditional probability
of the set of experiences $S'$ given the set of experiences $S$.
Still, I do not see any evidence that $f(S,S')$ should be
interpreted as a fundamental element of Mindless Sensationalism.
In any case, one can be conscious
only of a single conscious experience at once,
so there is no way in principle that one can test any properties
of joint sets of conscious experiences such as $f(S,S')$.

	An amusing property of both of these ad hoc ``conditional
probabilities'' for one conscious experience given another is that they
would both always be zero if the Orthogonal Projection Hypothesis were
true.  Even though the resulting theory would generally be a
``many-experiences'' theory, it could be interpreted as being rather
solipsistic in the sense that in the relative density matrix $\rho_p$
corresponding to my present conscious experience $p$, no other
disjoint set conscious experiences
would occur in it with nonzero measure!
This has the appearance of being somewhat unpalatable,
and might be taken to be an argument
against adopting the Orthogonal Projection Hypothesis,
but it is not clear to me that this is actually strong evidence
against the Orthogonal Projection Hypothesis.

	In addition to the fact that Mindless Sensationalism postulates
no fundamental notion of any {\it correlation} between individual
conscious experiences, it also postulates no fundamental {\it
equivalence} relation on the set of conscious experiences. For example,
the measure gives no way of classifying different conscious experiences
as to whether they belong to the same conscious being (e.g., at
different times) or to different conscious beings.  The most reasonable
such classification would seem to be by the content (including the {\it
qualia}) of the conscious experiences themselves, which distinguish the
conscious experiences, so that no two different conscious experiences,
$p\neq p'$, have the same content. Based upon my own present conscious
experience, I find it natural to suppose that conscious experiences
that could be put into the classification of being alert human
experiences have such enormous structure that they could easily
distinguish between all of the $10^{11}$ or so persons that are
typically assigned to our history of the human race.  In other words,
in practice, different people can presumably be distinguished by their
conscious experiences.

	Another classification of conscious experiences might be given
by classifying the awareness operators $A(S)$ rather than the content
of the conscious experiences themselves. This would be more analogous
to classifying people by the quantum nature of their bodies (in
particular, presumably by the characteristics of the relevant parts of
their brains).  However, I doubt that in a fundamental sense there is
any absolute classification that uniquely distinguishes each person in
all circumstances.  (Of course, one could presumably raise this
criticism about the classification of any physical object, such as a
``chair'' or even a ``proton'':  precisely what projection operators
correspond to the existence of a ``chair'' or of a ``proton''?) 
Therefore, in the present framework conscious experiences are
fundamental, but persons (or individual minds), like other physical
objects, are not, although they certainly do seem to be very good
approximate entities (perhaps as good as chairs or even protons) that I
do not wish to deny.  Even if there is no absolute definition of
persons in the framework of Mindless Sensationalism itself, the concept
of persons and minds does occur in some sense as part of the {\it
content} of my present conscious experience, just the concepts of
chairs and of protons do (in what are perhaps slightly different
``present conscious experiences,'' since I am not quite sure that I can
be consciously aware of all three concepts at once, though I seem to be
aware that I have been thinking of three concepts).

	In this way the framework of Mindless Sensationalism proposed
here is a particular manifestation of Hume's ideas \cite{Hume}, that
``what we call a {\it mind}, is nothing but a heap or collection of
different perceptions, united together by certain relations, and
suppos'd, tho' falsely, to be endow'd with a perfect simplicity and
identity'' (p. 207), and that the self is ``nothing but a bundle or
collection of different perceptions'' (p. 252).  As he explains in the
Appendix (p. 634), ``When I turn my reflexion on {\it myself}, I never
can perceive this {\it self} without some one or more perceptions; nor
can I ever perceive any thing but the perceptions.  'Tis the
composition of these, therefore, which forms the self.''  (Here I
should note that what Hume calls a perception may be only one {\it
component} of the ``phenomenal perspective'' or ``maximal experience''
\cite{Lo} that I have been calling a perception or conscious experience
$p$, so one of my $p$'s can include ``one or more perceptions'' in
Hume's sense.)

	Furthermore, each awareness operator $A(S)$ need not have any
precise location in either space or time associated with it, so there
need be no fundamental place or time connected with each conscious
experience. Indeed, Mindless Sensationalism can easily survive a
replacement of spacetime with some other structure (e.g., superstrings)
as more basic in the quantum world. Of course, the {\it contents} of a
conscious experience can include a sense or impression of the time of
the conscious experience, just as my present conscious experience when
I perceive that I am writing this includes a feeling that it is now
A.D. 2001, so the set of conscious experiences $p$ must include
conscious experiences with such beliefs, but there need not be any
precise time in the physical world associated with a conscious
experience.  That is, conscious experiences are `outside' physical
spacetime (even if spacetime is a fundamental element of the physical
world, which I doubt).

	As a consequence of these considerations, there are no unique
time-sequences of conscious experiences to form an individual mind or
self in Mindless Sensationalism. In this way the present framework
appears to differ from those proposed by Squires \cite{Sq}, Albert and
Loewer \cite{A,A92}, and Stapp \cite{St}.  (Stapp's also differs in
having the wavefunction collapse at each ``Heisenberg actual event,''
whereas the other two agree with mine in having a fixed quantum state,
in the Heisenberg picture, which never collapses.) Lockwood's proposal
\cite{Lo} seems to be more similar to mine, though he also proposes (p.
232) ``a continuous infinity of parallel such streams'' of
consciousness, ``{\it differentiating} over time,'' whereas Sensible
Quantum Mechanics has no such stream as fundamental.  On the other
hand, later Lockwood \cite{Lo2} does explicitly repudiate the
Albert-Loewer many-minds interpretation, so there seems to me to be
little disagreement between Lockwood's view and Mindless Sensationalism
except for the detailed formalism and manner of presentation.  Thus one
might label Mindless Sensationalism as the Hume-Everett-Lockwood-Page
(HELP) interpretation, though I do not wish to imply that these other
three scholars, on whose work my proposal is heavily based, would
necessarily agree with my present formulation,
which certainly is not contained in explicit detail
in what they have written.

	Of course, the conscious experiences themselves can include
components that seem to be memories of past conscious experiences or
events. In this way it can be a very good approximation to give an
approximate order for conscious experiences whose content include
memories that are correlated with the contents of other conscious
experiences. It might indeed be that the measure for conscious
experiences including detailed memories is rather heavily peaked around
approximate sequences constructed in this way.  But I would doubt that
the contents of the conscious experiences $p$, the awareness operators
$A(S)$, or the measures $\mu(S)$ for the sets of conscious experiences
$S$ would give unique sequences of conscious experiences that one could
rigorously identify with individual minds.

	Because the physical state of our universe seems to obey the
second law of thermodynamics, with growing correlations in some sense,
I suspect that the measure may have rather a smeared peak (or better,
ridge) along approximately tree-like structures of branching sequences
of conscious experiences, with conscious experiences further out along
the branches having contents that includes memories that are correlated
with the present-sensation components of conscious experiences further
back toward the trunks of the trees.  This is different from what one
might expect from a classical model with a discrete number of conscious
beings, each of which might be expected to have a unique sharp sequence
or non-branching trajectory of conscious experiences.  In the quantum
case, I would expect that what are crudely viewed as quantum choices
would cause smeared-out trajectories to branch into larger numbers of
smeared-out trajectories with the progression of what we call time.  If
each smeared-out trajectory is viewed as a different individual mind,
we do get roughly a ``many-minds'' picture that is analogous to the
``many-worlds'' interpretation \cite{E,DG}, but in my framework of
Mindless Sensationalism, the ``many minds'' are only approximate and
are not fundamental as they are in the proposal of Albert and Loewer
\cite{A}.  Instead, Mindless Sensationalism is a ``many-experiences''
or ``many-sensations'' interpretation.

	Even in a classical model, if there is one conscious experience
for each conscious being at each moment of time in which the being is
conscious, the fact that there may be many conscious beings, and many
conscious moments, can be said to lead to a ``many-experiences''
interpretation.  However, in Mindless Sensationalism, there may be
vastly more conscious experiences, since they are not limited to a
discrete set of one-parameter sharp sequences of conscious experiences,
but occur for all sets of conscious experiences $S$ for which $A(S)$ is
positive.  In this way a quantum model may be said to be even ``more
sensible'' (or is it ``more sensational''?) than a classical model. 
One might distinguish MS from a classical model with many conscious
experiences by calling MS a ``very-many-experiences'' framework,
meaning that almost all sets of possible conscious experiences actually
occur with nonzero measure. (Thus MS might, in a narrowly literal
sense, almost be a version of panpsychism, but the enormous range
possible for the logarithm of the measure means that it is really quite
far from the usual connotations ascribed to panpsychism.  This is
perhaps comparable to noting that there may be a nonzero amplitude that
almost any system, such as a star, has a personal computer in it,
and then calling the resulting many-worlds theory pancomputerism.)

	One might fear that the present attack on the assumption of any
definite notion of a precise identity for persons or minds as sequences
of conscious experiences would threaten human dignity.  Although I
would not deny that I feel that it might, I can point out that on the
other hand, the acceptance of the viewpoint of Mindless Sensationalism
might increase one's sense of identity with all other humans and other
conscious beings.  Furthermore, it might tend to undercut the
motivations toward selfishness that I perceive in myself if I could
realize in a deeply psychological way that what I normally anticipate
as my own future conscious experiences are in no fundamental way picked
out from the set of all conscious experiences.  (Of course, what I
normally think of as my own future conscious experiences are presumably
those that contain memory components that are correlated with the
content of my present conscious experience, but I do not see logically
why I should be much more concerned about trying to make such conscious
experiences happy than about trying to make conscious experiences happy
that do not have such memories:  better to do unto others as I would
wish they would do unto me.)  One can find that Parfit \cite{Par} had
earlier drawn similar, but much more sophisticated, conclusions from a
view in which a unique personal identity is not fundamental.

	The framework of Mindless Sensationalism can suggest various
questions, methods of analysis, and speculations that might not occur
to one using other frameworks.
I have done an analysis
\cite{SQM}
of the Einstein-Podolsky-Rosen (EPR) ``paradox''
\cite{EPR}
combined with that of Schr\"{o}dinger's cat
\cite{Sch},
finding that if the components of one's awareness are
correlated with different physical properties
that are highly correlated
(such as whether different parts of a cat are alive or dead),
then one can indeed predict that one's conscious experience will
have components that are highly correlated.
E.g., when one looks at the different parts of
Schr\"{o}dinger's cat, one will tend to have a strong agreement
between the components of the awareness of the different parts
of the cat's body as to whether the cat is dead or alive,
if indeed the actual awareness operators cause one to be 
aware of whether each part of the cat is dead or alive.
(If instead one were aware of whether each part of the cat were in
the symmetric or antisymmetric linear superposition
of being alive or dead, one would not have much agreement between
the components of the awareness of the separate parts
as to whether they were in the symmetric or antisymmetric states.)

	However, it still leaves it mysterious as to why we seem to be
aware of the properties that are highly correlated (such as whether the
different parts of a cat are dead or alive), rather than of properties
that are not highly correlated (such as whether the different parts of
a cat are in the symmetric or antisymmetric superpositions of being
dead or alive). In other words, it still is somewhat confusing to me
why in idealized cases our conscious experiences actually seem to be
rather unconfused. One might argue that if they were not unconfused,
then we could not act coherently and so would not survive. This would
seem to be a good argument only if our conscious experiences really do
affect our actions in the quantum world and are not just epiphenomena
that are determined by the quantum world without having any effect back
on it.  But on the other hand, it is not obvious how conscious
experiences could affect the quantum world in a relatively simple way
in detail (though it is easy to speculate on general ways in which
there might be some effect; see \cite{Page} and below).  So although it
appears to be unexplained, it conceivably could be that conscious
experiences do not affect the quantum world but are determined by it in
just such a way that in most cases they are not too confused. To mimic
Einstein, I am tempted to say, ``The most confusing thing about
conscious experiences is that they are generally unconfused.''

	As an aside, I should say that although epiphenomenalism seems
to leave it mysterious {\it why} typical conscious experiences are
unconfused, I do not think it leaves it mysterious {\it that} conscious
experiences occur, despite a na\"{\i}ve expectation that the latter is
also mysterious. The na\"{\i}ve argument is that if the conscious world
has no effect on the quantum world (usually called the physical world
\cite{Pen,Chal}, in contrast to my use of that term to include both the
quantum world and the conscious world), and if the development of life
in the quantum world occurs by natural selection, the development of
consciousness would have no effect on this natural selection and so
could not be explained by it.

	Nevertheless, one can give an answer analogous to what I have
heard was given by the late Fermilab Director Robert Wilson when he
was asked by a Congressional committee what Fermilab contributed to
the defense of the nation: ``Nothing.  But it helps make the nation
worth defending.'' Similarly, if epiphenomenalism is correct,
consciousness may contribute nothing to the survival of the species,
but it may help make certain species worth surviving. More accurately,
it may not contribute to the evolution of complexity, but it may select
us (probably not uniquely) as complex organisms which have typical
conscious experiences.  Then our consciousness would not be surprising,
because we are selected simply as typical conscious beings.

	This selection as typical conscious beings might also help
explain why we can do highly abstract theoretical mathematics and
physics that does not seem to help us much with our survival as a
species. If we are selected by the measure of our consciousness, and if
that is positively correlated with a certain kind of complexity that is
itself correlated with the ability to do theoretical mathematics and
physics, then it would not be surprising that we can do this better
than the average hominid that survives as well as we do (say averaging
over all the Everett many worlds).

	Another question one might ask
within the context of Mindless Sensationalism
is whether and how the measures of the sets of conscious experiences
associated with an individual brain depend on the brain
characteristics.  One might speculate that it might be greater for
brains that are in some sense more intelligent, so that in a crude
sense brighter brains have a bigger measure of conscious experiences. 
This could explain why you do not perceive yourself to be an insect,
for example, even though there are far more insects than humans.

	One might also be tempted to use this speculation to explain
why you may consider yourself to be more intelligent
than the average human
(though another possible explanation is that it is likely
that the average person considers himself brighter than average).
However, in this case the statistical
evidence, if present at all, is almost certainly much weaker
than in the case of comparing ourselves with ants.
Therefore, this speculation should not be used
to justify any politically incorrect conclusions that one
might be tempted to make from an assumption that he or she has
a greater measure of consciousness than most other humans.

	Also, one might conjecture that an appropriate measure on
conscious experiences might give a possible explanation of why most of
us perceive ourselves to be living on the same planet on which our
species developed.  This observation might seem surprising when one
considers that we may be technologically near the point at which we
could leave Earth and colonize large regions of the Galaxy \cite{Dys},
presumably greatly increasing the number of humans beyond the roughly
$10^{11}$ that are believed to have lived on Earth. If so, why don't we
have the conscious experiences of one of the vast numbers of humans
that may be born away from Earth? One answer is that some sort of doom
is likely to prevent this vast colonization of the Galaxy from
happening \cite{C,Le,N,G}, though these arguments are not conclusive
\cite{KKP}.  Although I would not be surprised if such a doom were
likely, I would na\"{\i}vely expect it to be not so overwhelmingly
probable that the probability of vast colonization would be so small as
is the presumably very small ratio of the total number of humans who
could ever live on Earth to those who could live throughout the Galaxy
if the colonization occurs.  Then, even though the colonization may be
unlikely, I would expect that it should still produce a higher measure
for conscious experiences of humans living off Earth than on it.

	However, another possibility is that colonization of the Galaxy
is not too improbable, but that it is mostly done by self-replicating
computers or machines who do not tolerate many humans going along, so
that the number of actual human colonizers is not nearly so large as
the total number who {\it could} live throughout the Galaxy if the
computers or machines did not dominate the colonization.  If the number
of these computers or machines dominate humans as ``intelligent''
beings (in the sense of having certain information-processing
capabilities), one might still have the question of why we perceive
ourselves as being humans rather than as being one of the vastly
greater numbers of such machines.  But the explanation might simply be
that the {\it measure} of conscious experiences is dominated by human
conscious experiences, even if the {\it number} of ``intelligent''
beings is not.  In other words, human brains may be much more efficient
in producing conscious experiences than the kinds of self-replicating
computers or machines which may be likely to dominate the colonization
of the Galaxy.  If such machines are more ``intelligent'' than humans
in terms of information-processing capabilities and yet are less
efficient in producing conscious experiences, our conscious experiences
of being human would suggest that the measure of conscious experiences
is not merely correlated with ``intelligence.''  (On the other hand, if
the measure of conscious experiences is indeed strongly correlated with
``intelligence'' in the sense of information-processing capabilities,
perhaps it might be the case that Galactic colonization is most
efficiently done by self-replicating computers or machines that are not
so ``intelligent'' as humans.  After all, insects and even bacteria
have been more efficient in colonizing a larger fraction of Earth than
have humans.)

	It might be tempting to take the observations that these
speculations might explain (our conscious experiences of ourselves as
human rather than as insect, and our experiences of ourselves as humans
on our home planet) as evidence tending to support the speculations. 
One could summarize such reasoning as a generalization of the Weak
Anthropic Principle \cite{D,Ca,CR,Ro,Da,BT,Les} that might be
called the {\it Conditional Aesthemic Principle}
(CAP, not entirely coincidentally the initials of my wife Cathy Anne):
given that we are conscious beings,
our conscious experiences are likely to be
typical experiences in the conscious world with its measure.

	Another use for the framework of Mindless Sensationalism
would be to see how various general approaches to the problems
of consciousness can be expressed in terms that are compatible
(in the way I have suggested) with quantum theory.
I have personally read so little of these approaches
(fewer books than I have fingers) that I am not competent
to try to see how to do that.
However, I must admit that from what little I have read
of, say functionalism, and from my mental attempts
to translate what I have read into the language
of my Mindless Sensationalism,
I am confused as to precisely how functionalism would be expressed.

	Functionalism is supposed to be
``the view that mental states are defined by their causes and effects''
\cite{Cam}.
If a particular ``mental state'' is to be identified
with a particular conscious experience $p$,
then I am not clear what its ``causes and effects''
are supposed to be.
Although I have no idea what the ``effects'' of $p$
are supposed to be,
I suppose that in one sense one could say that its causes are both
the experience operator $E(p)$
(the $p$-dependent operator whose sum or integral over the $p$'s
in the set $S$ gives the corresponding awareness operator $A(S)$)
and the quantum state of the universe, $\sigma$,
since both enter into the equation $m(p) = \sigma[E(p)]$
for the weight $m(p)$ that is summed or integrated over
the conscious experiences in a set $S$ to give the
measure $\mu(S)$ for that set.
If this interpretation of functionalism were correct,
a consequence for the conjecture of functionalism
would be that no two distinct conscious experiences,
say $p$ and $p'$, have the same experience operators:
If $p \neq p'$, then $E(p) \neq E(p')$.
Equivalently, if $E(p) = E(p')$, then $p = p'$.
This is certainly a plausible conjecture,
but I see no way to justify it or test whether or not
it is true, though I believe that it is a conjecture
with real content and logically could be either true or false.

	Another interpretation might be to identify
a ``mental state'' with a quantum state that
gives rise to a particular conscious experience $p$.
If any state $\sigma$ that gives $m(p) = \sigma[E(p)] > 0$
is counted as a ``mental state'' that ``gives rise''
to $p$, then all but a set of measure zero of possible
quantum states $\sigma$ could be said to ``give rise'' to $p$.
This seems far too broad, so let us see whether we can
get a narrower class of quantum states that ``give rise'' to $p$.

	One way is to consider what different quantum states
can be considered to contribute ``directly''
to a conscious experience $p$.
If, for a given conscious experience $p$,
the corresponding experience operator $E(p)$
were decomposed into a weighted sum
of orthogonal rank-one projection operators $\Pi_i$,
 \begin{equation}
 E(p) = \sum_i W_i \Pi_i
 \label{eq:13}
 \end{equation}
with positive weights $W_i$,
then the eigenstate $|\psi_i\rangle$ with unit eigenvalue
of each of these projection operators $\Pi_i$
(the state which when written in the form $|\psi_i\rangle\langle\psi_i|$
is identical to the rank-one projection operator $\Pi_i$)
would give a contribution to the measure for the conscious
experience $p$.
In a sense one can say that it is each of these eigenstates
(one for each rank-one projection operator
that occurs in Eq.~(\ref{eq:13}))
that {\it directly} gives rise to the conscious experience $p$.
(Of course, any state $\sigma$ that is not orthogonal to all of these
eigenstates will give a positive weight
for the the conscious experience $p$,
 \begin{equation}
 m(p) = \langle E(p)\rangle = \sum_i W_i \sigma[\Pi_i]
 \label{eq:14}
 \end{equation}
the weighted sum of the overlaps of the state $\sigma$
with the eigenstates $\Pi_i = |\psi_i\rangle\langle\psi_i|$.
But it is the eigenstates themselves that can be considered
to be most directly related to the conscious experience $p$.)

	So if the ``mental states'' corresponding to
the conscious experience $p$ are defined to be the eigenstates
$\Pi_i$ that occur in the sum given by Eq.~(\ref{eq:13})),
the we can ask what the ``causes and effects'' of these are.
If an answer to that could be found, perhaps the conjecture
of functionalism might be that any two ``mental states'' $\Pi_i$
corresponding to the same conscious experience $p$
would have the same ``causes and effects.''
Or it might be the converse, that for any ``mental state'' $\Pi_i$
that occurs in the sum given by Eq.~(\ref{eq:13})),
any other rank-one projection
operator with the same ``causes and effects'' also occur in that sum.
Either of these two conjectures seems to have nontrivial content,
but precisely what that content would be depends upon
what ``having the same causes and effects'' is taken to mean.
Without an understanding of that, my attempt to guess precisely
what functionalism might mean remain stymied.

	Therefore, it would be interesting indeed
to see how functionalism might possibly be expressed
in terms of the operators $E(p)$ and $A(S)$
that occur in Mindless Sensationalism.

	I have used the example of functionalism
not merely to express my own confusion
(which might be merely due to my gross ignorance of the field),
but also to illustrate that if one can translate
conjectures from the philosophy of mind into
the language of Mindless Sensationalism,
one may be able to come up with some precise formulations for them
that would be applicable to the real universe
and not just to some imaginary universe that is
modeled by, say, some classical Turing machine.

	Similarly, it would also be an interesting challenge to
interpret other approaches to the problems of consciousness
within the framework of Mindless Sensationalism.
If they cannot be interpreted within this framework,
one would need to invent another framework in which
they might be interpreted in order for them to be consistent
with our quantum universe.
This might impose a nontrivial constraint on
approaches to the problems of consciousness.

	In conclusion, I am proposing that Mindless Sensationalism is
the best framework we have at the present level for understanding the
connection between conscious experiences and quantum theory.  Of
course, the framework would only become a complete theory once one had
the set $M$ of all conscious experiences $p$, the awareness operators
$A(S)$, and the quantum state $\sigma$ of the universe.

	Even such a complete theory of the quantum world and the
conscious world affected by it need not be the ultimate simplest
complete theory of the combined physical world.  There might be a
simpler set of unifying principles from which one could in principle
deduce the conscious experiences, awareness operators, and quantum
state, or perhaps some simpler entities that replaced them.  For
example, although in the present framework of Mindless Sensationalism,
the quantum world (i.e., its state), along with the awareness
operators, determines the measure for experiences in the conscious
world, there might be a reverse effect of the conscious world affecting
the quantum world to give a simpler explanation than we have at present
of the coherence of our conscious experiences and of the correlation
between will and action (why my desire to do something I feel am
capable of doing is correlated with my conscious experience of actually
doing it, i.e., why I ``do as I please'').  If the quantum state is
partially determined by an action functional, can desires in the
conscious world affect that functional (say in a coordinate-invariant
way that therefore does not violate energy-momentum conservation)? 
Such considerations may call for a more unified framework than Mindless
Sensationalism (elsewhere called Sensible Quantum Mechanics), which one
might call Sensational Quantum Mechanics \cite{SQM,Page}. Such a more
unified framework need not violate the limited assumptions of Mindless
Sensationalism, though it might do that as well and perhaps reduce to
Mindless Sensationalism only in a certain approximate sense.

	To explain these frameworks in terms of an analogy, consider a
classical model of spinless massive point charged particles and an
electromagnetic field in Minkowski spacetime.  Let the charged
particles be analogous to the quantum world (or the quantum state part
of it), and the electromagnetic field be analogous to the conscious
world (the set of conscious experiences with its measure $\mu(S)$).  At
the level of a simplistic materialist mind-body philosophy, one might
merely say that the electromagnetic field is part of, or perhaps a
property of, the material particles.  At the level of Mindless
Sensationalism, the charged particle worldlines are the analogue of the
quantum state, the retarded electromagnetic field propagator (Coulomb's
law in the nonrelativistic approximation) is the analogue of the
awareness operators, and the electromagnetic field determined by the
worldlines of the charged particles and by the retarded propagator is
the analogue of the conscious world.  (Here one can see that this
analogue of Mindless Sensationalism is valid only if there is no free
incoming electromagnetic radiation.)  At the level of Sensational
Quantum Mechanics, at which the conscious world may affect the quantum
world, the charged particle worldlines are partially determined by the
electromagnetic field through the electromagnetic forces that it
causes.  (This more unified framework better explains the previous
level but does not violate its description, which simply had the
particle worldlines given.)  At a yet higher level, there is the
possibility of incoming free electromagnetic waves, which would violate
the previous frameworks that assumed the electromagnetic field was
uniquely determined by the charged particle worldlines.  (An analogous
suggestion for intrinsic degrees of freedom for consciousness has been
made by the physicist Andrei Linde \cite{Lin90}.)  Finally, at a still
higher level, there might be an even more unifying framework in which
both charged particles and the electromagnetic field are seen as modes
of a single entity (e.g., to take a popular current speculation, a
superstring, or perhaps some more basic entity in ``M theory'').

	Therefore, although it is doubtful that Mindless Sensationalism
is the correct framework for the final unifying theory (if one does
indeed exist), it seems to me to be a move in that direction that is
consistent with what we presently know about the physical world and
consciousness.

	This work has been supported in part by
the Natural Sciences and Engineering Council of Canada.
Many of the people whom I have remembered as being influential
in my formulation of my ideas are listed at the end of
\cite{SQM},
though of course none of them are ultimately responsible for it,
and indeed most of them might well disagree with it.


\baselineskip 7pt

\begin{thebibliography}{99}

\bibitem{HarHaw} J. B. Hartle and S. W. Hawking,
Phys.\ Rev.\ D{\bf 28}, 2960 (1983).

\bibitem{Smith} Q. Smith,
``Why cognitive Scientists Cannot Ignore Quantum Mechanics,''
to be published in {\em Consciousness:  New Philosophical Essays},
edited by Quentin Smith and Alexandar Jokic
(Oxford: Oxford University Press, 2002).

\bibitem{Loewer} B. Loewer,
``Consciousness and Quantum theory,''
to be published in {\em Consciousness:  New Philosophical Essays},
edited by Quentin Smith and Alexandar Jokic
(Oxford: Oxford University Press, 2002).

\bibitem{Lockwood} M. Lockwood,
``Consciousness and the Quantum World:
Putting Qualia on the Map,''
to be published in {\em Consciousness:  New Philosophical Essays},
edited by Quentin Smith and Alexandar Jokic
(Oxford: Oxford University Press, 2002).

\bibitem{SQM}  D. N. Page,
``Sensible Quantum Mechanics:
Are Only Perceptions Probabilistic?''
(University of Alberta report Alberta-Thy-05-95, 1995 June 7;
revised 1997 June 30),
quant-ph/9506010.

\bibitem{Page} D. N. Page, ``Probabilities Don't Matter,''
in {\em Proceedings of the 7th Marcel Grossmann Meeting
on General Relativity},
edited by R. T. Jantzen and G. M. Keiser
(World Scientific, Singapore 1996),
pp. 983-1002, gr-qc/9411004;
``Information Loss in Black Holes and/or Conscious Beings?''
in {\em Heat Kernel Techniques and Quantum Gravity},
edited by S. A. Fulling
(Discourses in Mathematics and Its Applications,
No. 4, Texas A\&M University Department of Mathematics,
College Station, Texas, 1995), pp. 461-471, hep-th/9411193;
``Attaching Theories of Consciousness to Bohmian Quantum Mechanics,''
in {\em Bohmian Quantum Mechanics and Quantum Theory:  An Appraisal},
edited by J. T. Cushing, A. Fine, and S. Goldstein
(Kluwer, Dordrecht, 1996), pp. 197-210, quant-ph/9507006;
Int. J. Mod. Phys. {\bf D5}, 583-596 (1996), gr-qc/9507024;
``Aspects of Quantum Cosmology,''
in {\em String Gravity and Physics at the Planck Energy Scale}
(NATO ASI Series, Series C: Mathematical and Physical Sciences
 -- Vol 476;
Proceedings of the International School of Astrophysics
``D. Chalonge,'' 4th Course, Erice, Sicily, 8-19 September 1995),
edited by N. Sanchez and A. Zichichi
(Kluwer, Dordrecht, 1996), pp. 431-450, gr-qc/9507025;
``Quantum Cosmology Lectures,''
in {\em Proceedings of the First Mexican School
on Gravitation and Mathematical Physics,
Guanajuato, Mexico, Dec. 12-16, 1994},
edited by A. Macias, T. Matos, O. Obregon, and H. Quevedo
(World Scientific, Singapore, 1996), pp. 70-86, gr-qc/9507028.

\bibitem{Lo} M. Lockwood, {\em Mind, Brain and the Quantum:
The Compound `I'}
(Basil Blackwell, Oxford, 1989).

\bibitem{Lo2} M. Lockwood, in {\em Erwin Schr\"{o}dinger:  Philosophy
and the Birth of Quantum Mechanics}, edited by M. Bitbol and O.
Darrigol (Editions Fronti\`{e}res, Gif-sur-Yvette Cedex, 1992), p.~363.

\bibitem{Lo96} M. Lockwood,
`` `Many Minds' Interpretations of Quantum Mechanics,''
Brit. J. Phil. Sci. {\bf 47}, 445-461 (1996).

\bibitem{A} D. Albert and B. Loewer,
Synthese {\bf 77}, 195 (1988);
{\bf 86}, 87 (1991).

\bibitem{A92} D. Z. Albert, {\em Quantum Mechanics and Experience}
(Harvard University Press, Cambridge, Massachusetts, 1992).

\bibitem{Chal} D. J. Chalmers,
{\em The Conscious Mind:  In Search of a Fundamental Theory}
(Oxford University Press, New York, 1996).

\bibitem{Dav} E. B. Davies, {\em Quantum Theory of Open Systems}
(Academic Press, London, 1976); I thank Shelly Goldstein (private
communication) for suggesting the use of POV measures and directing me
to this reference.	

\bibitem{deB} L. de Broglie,
``La nouvelle dynamique des quanta,''
in {\em Electrons et Photons: Rapports et Discussions du Cinqui\`eme
Conseil de Physique tenu \`a Bruxelles du 24 au 29 Octobre 1927
sous les Auspices de l'Institut International de Physique Solvay}
(Gauthier-Villars, Paris, 1928), pp. 105-132;
{\em Physicien et Penseur}  
(Paris, 1953), p. 465;
{\em Tentative d'Interpr\'{e}tation Causale et
Non-lin\'{e}aire de la M\'{e}canique Ondulatoire}.
(Gauthier-Villars, Paris, 1956).
{\em Foundations of Physics} {\bf 1}, 5 (1970).

\bibitem{Bohm} D. Bohm, Phys.\ Rev.\ {\bf 85}, 166-179, 180-193 (1952);
 Phys.\ Rev.\ {\bf 89}, 458-466 (1953);
D. Bohm and B. J. Hiley,
{\em The Undivided Universe:  An
Ontological Interpretation of Quantum Theory}.
(Routledge and Kegan Paul, London, 1993).

\bibitem{Hol} P. R. Holland, {\em The Quantum Theory of Motion}
(Cambridge University Press, Cambridge, 1993).

\bibitem{BDDGZ}
K. Berndl, M. Daumer, D. Durr, S. Goldstein, and N. Zanghi,
Nuovo Cim.\ {\bf B110}, 737-750 (1995), quant-ph/9504010.

\bibitem{CFG} J. T. Cushing, A. Fine, and S. Goldstein, eds.,
{\em Bohmian Quantum Mechanics and Quantum Theory:  An Appraisal},
(Kluwer, Dordrecht, 1996).

\bibitem{Sid93} S. Coleman, ``Quantum Mechanics with the Gloves Off,''
Dirac Memorial Lecture, St. John's College, University of Cambridge,
June 1993 (unpublished).

\bibitem{Sid95} S. Coleman, Physics Colloquium, University of Alberta,
March 31, 1995 (unpublished).

\bibitem{E} H. Everett, III, Rev.\ Mod.\ Phys.\ {\bf 29}, 454 (1957).

\bibitem{DG} B. S. DeWitt and N. Graham, eds., {\em The Many-Worlds
Interpretation of Quantum Mechanics} (Princeton University Press,
Princeton, 1973).

\bibitem{Sel} W. Sellars, J.\ Metaphysics {\bf 18}, 430-451 (1965).

\bibitem{Bar} J. Barbour, {\em The End of Time:
The Next Revolution in Our Understanding of the Universe}
(Weidenfeld \& Nicolson, London, 1999).

\bibitem{Hume} D. Hume, {\em A Treatise of Human Nature}, reprinted
from the original edition in three volumes and edited by L. A.
Selby-Bigge (Clarendon, Oxford, 1888).

\bibitem{Sq} E. J. Squires, Found.\ Phys.\ Lett.\ {\bf 1}, 13 ((1987);
{\em Conscious Mind in the Physical World}
(Adam Hilger, Bristol and New York, 1990);
Synthese {\bf 89}, 283 (1991); {\bf 97}, 109 (1993).

\bibitem{St} H. P. Stapp, {\em Mind, Matter, and Quantum Mechanics}
(Springer-Verlag, Berlin, 1993);
``The Integration of Mind into Physics''
(Lawrence Berkeley Laboratory report LBL-35880, July 13, 1994);
``Is Mental Process Noncomputable?''
(Lawrence Berkeley Laboratory report LBL-36345, Dec. 1994),
quant-ph/9502011;
``Why Classical Mechanics Cannot Naturally Accommodate Consciousness
But Quantum Mechanics Can''
(Lawrence Berkeley Laboratory report LBL-36574, Feb. 8, 1995),
quant-ph/9502012;
``Quantum Mechanical Coherence, Resonance, and Mind''
(Lawrence Berkeley Laboratory report LBL-36915, Nov. 1994),
quant-ph/9504003;
``The Hard Problem:  A Quantum Approach''
(Lawrence Berkeley Laboratory report LBL-37163, May 1995),
quant-ph/9505023;
``Values and the Quantum Concept of Man''
(Lawrence Berkeley Laboratory report LBL-37315, June 1995),
quant-ph/950603;
``Chance, Choice, and Consciousness:
The Role of Mind in the Quantum Brain''
(Lawrence Berkeley Laboratory report LBL-37944, Nov. 1995),
quant-ph/9511029;
``Science of Consciousness and the Hard Problem''
(Lawrence Berkeley Laboratory report LBL-38621, Apr. 1996);
``Nonlocal Character of Quantum Theory''
(Lawrence Berkeley Laboratory report LBL-38803, May 1996);
``Review of Chalmer's Book,''
(Lawrence Berkeley Laboratory report LBL-38890, May 1996);
``The Evolution of Consciousness''
(Lawrence Berkeley Laboratory report LBL-39241, Aug. 1996);
``Quantum Ontologies and Mind-Matter Synthesis,''
quant-ph/9905053;
``Whiteheadian Process and Quantum Theory of Mind,''
(Lawrence Berkeley Laboratory report LBL-42143, Aug. 1998),
``Attention, Intention, and Will in Quantum Physics,''
(Lawrence Berkeley Laboratory report LBL-42650, May 1999),
quant-ph/9905054.

\bibitem{Par} D. Parfit, Phil.\ Rev.\ {\bf 80} (1971),
reprinted in {\em Personal Identity}, edited by J. Perry
(University of California Press, Berkeley, 1975), p.~199;
{\em Reasons and Persons} (Clarendon Press, Oxford, 1984).
I thank M. Lockwood (private communication)
for drawing these references to my attention.

\bibitem{EPR} A. Einstein, B. Podolsky, and N. Rosen,
Phys.\ Rev.\ {\bf 47}, 777 (1935),
reprinted in {\em Quantum Theory and Measurement},
edited by J. A. Wheeler and W. H. Zurek
(Princeton University Press, Princeton, 1983), p.~138.

\bibitem{Sch} E. Schr\"{o}dinger, Naturwissenschaften
{\bf 23}, 807 (1935),
English translation by J. D. Trimmer,
Proc. Am.\ Phil.\ Soc.\ {\bf 124}, 323 (1980),
reprinted in {\em Quantum Theory and Measurement},
edited by J. A. Wheeler and W. H. Zurek
(Princeton University Press, Princeton, 1983), p.~152.

\bibitem{Pen} R. Penrose,
{\em Shadows of the Mind:
A Search for the Missing Science of Consciousness}
(Oxford University Press, Oxford, 1994).

\bibitem{Dys} F. Dyson, {\em Imagined Worlds}
(Harvard University Press, Cambridge, 1997).

\bibitem{C} B. Carter,
Phil.\ Trans.\ Roy.\ Soc.\ Lond.\ {\bf A310}, 347
(1983).

\bibitem{Le} J. Leslie, Bull. Canad. Nucl. Society {\bf 10}, 10 (1989);
Interchange {\bf 21}, 49-58 (1990);
{\em Universes} (Routledge, London and New York, 1989), p.~214;
Phil. Quart. {\bf 40}, 65 (1990); {\bf 42}, 85 (1992);
Mind {\bf 101}, 521 (1992); {\bf 102}, 489 (1993);
Math.\ Intelligencer {\bf 14}, 48 (1992); {\bf 15}, 5 (1993);
Interchange {\bf 23}, 289 (1992);
J.\ Appl.\ Phil. {\bf 11}, 31 (1994);
{\em The End of the World:  The Science and Ethics of Human Extinction}
(Routledge, London and New York, 1996).

\bibitem{N} H. B. Nielsen, Acta Physica Polonica {\bf B20}, 427 (1989).

\bibitem{G} J. R. Gott III, Nature {\bf 363}, 315 (1993); {\bf 368}, 108
(1994).

\bibitem{KKP} T. Kopf, P. Krtou\v{s}, and D. N. Page,
``Too Soon for Doom Gloom?''
(University of Alberta report Alberta-Thy-17-94, 1994 July 4),
gr-qc/9407002.

\bibitem{D} R. H. Dicke,
Rev.\ Mod.\ Phys.\ {\bf 29}, 355 and 363 (1977);
Nature {\bf 192} 440 (1961).

\bibitem{Ca} B. Carter,
in {\em Confrontation of Cosmological Theories with Observation},
edited by M. S. Longair (Reidel, Dordrecht, 1974), p.~291.

\bibitem{CR} B. J. Carr and M. J. Rees, Nature {\bf 278}, 605 (1979).

\bibitem{Ro} I. L. Rozental, Sov.\ Phys.\ Usp.\ {\bf 23}, 296 (1980).

\bibitem{Da} P. C. W. Davies,
{\em The Accidental Universe}
(Cambridge University Press, Cambridge, 1982).

\bibitem{BT} J. D. Barrow and F. T. Tipler,
{\em the Anthropic Cosmological Principle}
(Clarendon Press, Oxford, 1986).

\bibitem{Les} J. Leslie, Am.\ Phil.\ Quart., 141 (April 1982);
Mind, 573 (October 1983);
in {\em Current Issues in Teleology}, edited by N. Rescher
(University Press of America, Lanham and London, 1983), p.~111;
in {\em Proceedings of the Philosophy of Science Association 1986}
(Edwards Bros, Ann Arbor, 1986), vol.~1, p.~87;
in {\em Origin and Early History of the Universe},
edited by J. Demaret (University of Li\`{e}ge, Li\`{e}ge, 1987), p.~439;
Mind, 269 (April 1988);
{\em Universes} (Routledge, London and New York, 1989);
{\em Physical Cosmology and Philosophy} (Macmillan, New York, 1990).

\bibitem{Cam} R. Audi, ed.,
{\em The Cambridge Dictionary of Philosophy}
(Cambridge University Press, Cambridge, 1995), p. 288.

\bibitem{Lin90} A. Linde,
{\it Particle Physics and Inflationary Cosmology}
(Harwood Academic Publishers, Chur, Switzerland, 1990), p.~317.

\end{thebibliography}
\end{document}